\input jnl

\def\gcall#1{(\call{#1})}

\def\vvec#1{ {\bf #1} }
\def\L{{\cal L}}

\def\F{{\cal F}}

\def\O{{\cal O}}

\head{\bf HIGH FRICTION LIMIT OF THE KRAMERS EQUATION : THE MULTIPLE TIME 
SCALE APPROACH }

\author Lyd\'eric Bocquet 

\affil Laboratoire de Physique, Ecole Normale Sup\'erieure
de Lyon (URA CNRS 1325), 
46 All\'ee d'Italie, 69007  Lyon, France.
e-mail : lbocquet@physique.ens-lyon.fr

\abstract
The purpose of the paper is to give a pedagogical introduction to
the multiple time-scale technique, on the example of the high
friction limit of the Kramers equation. We begin with a discussion of
the standard perturbation technique as presented in van Kampen's reference
book \refto{VK}, which will be shown to fail in the long-time limit.
Application of the multiple time-scale technique avoids these difficulties
and leads to a uniform expansion in powers of the inverse of the friction.
Analogy with the Chapman-Enskog expansion is discussed.

 Accepted for publication in American Journal of Physics 

\smallskip 

\dateline

\endpage

\noindent
{\bf I. INTRODUCTION}
\taghead{I.}

From a pedagogical point of view, the theory of Brownian motion is of 
fundamental interest in a graduate course on Statistical Mechanics. 
Starting with quite simple assumptions, a microscopic stochastic model 
- namely the Langevin model - can be introduced to describe the thermal motion
of the suspended Brownian particles. 
It is noteworthy that this simple treatment contains already
all the basic features of non-equilibrium thermodynamics,
such as diffusive motion, fluctuation-dissipation theorem, Einstein relation between mobility and diffusion and the Green-Kubo relation for the diffusion 
coefficient \refto{VK}. A complete description requires however a more precise 
characterization of the stochastic processes involved in Brownian motion,
culminating in the Fokker-Planck evolution equation
for the probability distribution of the Brownian particle \refto{VK,DGM}.
Two levels of description can be used. First, if both the velocity and 
position of the particle are considered, the probability distribution
$P(\vvec r,\vvec v,t)$
evolves according to the Kramers equation 
$$
 \Biggl[{\partial \over \partial t}+{\vvec v}.\vvec{\nabla}_{\vvec r}+
 {\vvec\F(\vvec r) \over M}.\vvec{\nabla}_{\vvec v}\Biggr] P(\vvec r,\vvec v,t)
 = \xi \vvec{\nabla}_{\vvec v}.\Biggl\{\biggl(\vvec{v} +{k_BT\over
 M}\vvec{\nabla}_{\vvec v} \biggr) P(\vvec r,\vvec v,t) \Biggr\}
 \tag{1}
 $$
where $M$ denotes the mass of the Brownian particle, $\xi$ the friction, $k_BT$ the bath 
temperature and $\vvec\F(\vvec r)$ an external force field.
This description was first given by Klein \refto{Klein21} and 
Kramers \refto{Kramers40}. On the other hand, if only the spatial evolution of the particle is considered, one may rather use the Smoluchowski equation,
governing the time evolution of the probability density $\rho(\vvec r,t)$
in configuration space :
$$
{\partial \over \partial t} \rho (\vvec r,t)
 = {1 \over {M\xi}} \vvec{\nabla}_{\vvec r}.\Biggl\{ k_BT \vvec{\nabla}_{\vvec r} -
 \vvec\F(\vvec r) \Biggr\} \rho (\vvec r,t)
 \tag{2}
 $$
This description, first put forward by Smoluchowski \refto{Smolu16}, 
predates
that of Kramers and Klein. 

However it took a long time to clarify the link between both descriptions. Kramers \refto{Kramers40} already pointed out that the Smoluchowski
description should be obtained from the Kramers equation in the high-friction
limit. This can indeed be intuitively understood : if
the friction is high, the velocity relaxes after a short time, of order $1/\xi$,
and the evolution on larger time scales is only determined by the spatial 
distribution. However, a rigorous perturbative derivation of the Smoluchowski
equation from the Kramers one was given by Wilemski only in 1976 \refto{Wilemski76}. Another systematic expansion procedure was later suggested by Titulaer \refto{Titulaer78}. The latter is closely related to the
Chapman-Enskog perturbation scheme to solve the Boltzmann equation\refto{Cercignani}. This method is indeed natural in the present case,
in view of the formal analogy between the collisional terms occuring in the
Boltzmann and in the Kramers equations. In both equations, the right-hand
side characterizes the effect of collisions on the time evolution of the
distribution function.
The singular nature of the
expansion is due to the presence of the small parameter
$1/\xi$ (replaced by
the Knudsen number in the case of the Boltzmann equation)
multiplying the time derivative in the Kramers equation. This makes a straightforward application of the perturbation
theory impossible.
In his reference textbook \refto{VK} (pp. 234-236), van
Kampen proposes an alternative, pedagogical method, which is closer in 
spirit to 
the Hilbert perturbation scheme \refto{Cercignani}. This procedure
leads indeed to the Smoluchowski equation. The purpose of this paper is
to show that this pedagogical approach is however incomplete and leads to
inconsistencies in the expansion procedure.
In section II, we will show that hidden secular terms are present in this
description, which
diverge as time goes to infinity. These terms make the asymptotic expansion non-uniform in the small parameter $1/\xi$. In section
III, we propose an 
alternative approach, based on the multiple-time scale method. This procedure eliminates
the secular divergences and construct a uniform expansion in the small
parameter (the inverse friction in our case) \refto{Nayfeh}.
The multiple time-scale method allows one to recover quite easily all the features of the Chapman-Enskog
procedure of Titulaer \refto{Titulaer78}. 
Moreover, it gives a clear interpretation of the contribution of
all orders in $1/\xi$ in the time derivative, as it is formally introduced
in the Chapman-Enskog {expansion\refto{Cercignani}}.

\noindent
{\bf II. THE HILBERT EXPANSION}
\taghead{II.}

In this section, we first reproduce and then discuss the expansion procedure as given
in van Kampen's reference book, in order to point out the limitation of this ``Hilbert''
perturbation scheme. We will restrict 
ourselves to the one-dimensional case to simplify the analysis. 

First, we introduce new dimensionless variables 
$$
\tau=t\ {v_T\over \ell};\ \ \  V={v\over v_T} ;\ \ \  X={x\over \ell} ;\ \ \ F=\F{\ell\over Mv_T^2}
\tag{3 .a}
 $$
and a dimensionless friction
$$
\xi_d=\xi {\ell\over v_T}
\tag{3 .b}
$$
where $v_T=\sqrt{k_BT/M}$ is the thermal velocity and $\ell$ is a characteristic
length scale of the system (such as the Brownian particle diameter).
The Kramers equation then takes the following form
 $$
 {\partial \over \partial V}\Biggl(V+{\partial \over \partial V}\Biggr) P(X,V;\tau)
 = {1\over \xi_d} \Biggl[{\partial \over \partial \tau}+V{\partial\over\partial X}
 +F(X){\partial \over \partial V}\Biggr] P(X,V;\tau)
 \tag{4}
 $$
Let us insert into \gcall{4} the ``naive'' expansion
$$
P=P^{(0)}~+~\xi_d^{-1}~P^{(1)}~+~\xi_d^{-2}~P^{(2)}~+~\dots
\tag{5}
$$
By identifying terms of the same order, we find the following equations
$$
\eqalign{
&\L_{FP}\ P^{(0)}\ = \ 0 \cr
&\L_{FP}\ P^{(1)}\ = \ \Biggl[{\partial \over \partial \tau}+V{\partial\over\partial X}
 +F(X){\partial \over \partial V}\Biggr] P^{(0)} \cr
&\L_{FP}\ P^{(2)}\ = \ \Biggl[{\partial \over \partial t}+V{\partial\over\partial X}
 +F(X){\partial \over \partial V}\Biggr] P^{(1)} \cr
& \dots \cr
}
\tag{6}
$$
where we introduced a ``Fokker-Planck'' operator $\L_{FP}$ defined as
$$
\L_{FP}={\partial \over \partial V}\Biggl(V+{\partial \over \partial V}\Biggr)
\tag{7}
$$
The zeroth order equation imposes a maxwellian velocity distribution
$$
P^{(0)}(X,V;\tau)=\Phi (X;\tau) e^{-1/2 V^2}
\tag{8}
$$
and the function $\Phi$ has to be determined.

The first order term then
gives
$$
\L_{FP}\ P^{(1)}\ = \ {\partial \Phi \over \partial \tau} e^{-1/2 V^2}+ V\left\{{\partial \Phi \over\partial X}
 -F\ \Phi \right\} e^{-1/2 V^2}
\tag{9}
$$
By integrating both sides over $V$, one obtains a ``solubility condition''
$$
{\partial \Phi \over \partial \tau}=0
\tag{10}
$$
This can be understood in terms of linear algebra. The eigenfunctions of $\L_{FP}$ are the functions
$H_n(V/\sqrt{2})\exp(-V^2/2)$, where $H_n$ is the $n$-th Hermite polynomial. Since
the Maxwellian is associated with a null eigenvalue, the orthogonality
between the eigenfunctions imposes that all terms multiplying the
maxwellian on the right hand-side (r.h.s.) of equations like \gcall{9} must be vanish. This is
precisely the ``solubility condition'' \gcall{10}.

Equation \gcall{9} can now be solved to yield
$$
P^{(1)}(X,V;\tau)=-V\left\{{\partial \Phi \over\partial X}
 -F\ \Phi \right\} e^{-1/2 V^2}+\Psi(X;\tau) e^{-1/2 V^2}
\tag{11}
$$
where $\Psi$ has to be determined.

By inserting this solution into \gcall{6}, one obtains 
$$
\eqalign{
\L_{FP}\ P^{(2)}\ =\ &\left\{ {\partial \Psi \over \partial \tau} 
- {\partial \over \partial X } \left( {\partial \Phi \over\partial X}
 -F\ \Phi\right) \right\} e^{-1/2 V^2} + \left\{{\partial \Psi \over\partial X} -F\ \Psi \right\} V\ e^{-1/2 V^2} \cr
&-\left\{ 
F\left( {\partial \Phi \over\partial X}
 -F\ \Phi \right) -{\partial \over \partial X } \left( {\partial \Phi \over\partial X}
 -F\ \Phi\right) 
\right\} (1-V^2)\ e^{-1/2 V^2}
}
\tag{12}
$$
where the terms have been grouped into Hermite polynomials $H_n(V/\sqrt{2})$.
According to the solubility condition, the first term in the
r.h.s. of \gcall{12} must be set to zero, so that
$$
{\partial \Psi \over \partial \tau} 
- {\partial \over \partial X } \left( {\partial \Phi \over\partial X}
 -F\ \Phi\right) = 0
\tag{13}
$$
Following van Kampen, one can then collect the results to
obtain the distribution function to order $1/\xi_d^{2}$ :
$$
P(X,V;\tau)=e^{-1/2 V^2} \left[ \Phi (X) - \xi_d^{-1} V\ \left(
{\partial \Phi \over\partial X} -F\ \Phi \right)+ \xi_d^{-1} \Psi(X;\tau)
+ \O (\xi_d^{-2}) \right]
\tag{14}
$$
After integrating over the velocity, one obtains the spatial probability density  $\rho(X;\tau)$ as
$$
\rho(X;\tau)=\sqrt{2\pi}\left[ \Phi(X) + \xi_d^{-1} \Psi(X;\tau)+ \O (\xi_d^{-2}) \right]
\tag{15}
$$
and eq. \gcall{13} then yields the (dimensionless) Smoluchowski equation
$$
{\partial \rho(X,\tau) \over \partial \tau} 
= {1 \over \xi_d} {\partial \over \partial X } \left( {\partial \rho(X,\tau) \over\partial X}
 -F(X)\ \rho(X,\tau)\right) 
\tag{16}
$$
However, a careful inspection of eq. \gcall{13} shows that $\Psi$ 
diverges as time goes to infinity. Indeed, eq. \gcall{10} shows that
$\Phi$ is independent of time. Then, the first term in \gcall{13} does not
depend on time either and the function $\Psi$ is proportionnal 
to time $\tau$
$$
\Psi(X;\tau) \sim \tau
\tag{17}
$$
The standard expansion \gcall{5} of the solution thus leads to 
secular divergences and cannot be correct in the long time limit.
In other words, a naive expansion is not uniformly convergent for
small $1/\xi_d$. The limit of time going to infinity cannot be inverted with the limit of friction going to infinity, so that 
taking small corrections in $1/\xi_d$ into account leads to an upper bound
for the time $\tau$. In fact, this procedure implicitly expands time
dependent terms
like $\tau/\xi_d$ as $\xi_d^{-1}$ terms, and secular divergences appear in the 
long-time limit.

\noindent
{\bf III. THE MULTIPLE TIME-SCALE APPROACH}
\taghead{III.}

In the limit $\xi^{-1} \rightarrow 0$, the Kramers argument presented above
indicates a  time-scale separation. There is first a very short period ($t \sim \xi^{-1}$) during
which the velocity of the Brownian particle thermalizes. Then the dynamical
evolution is controlled by the time-dependence of the spatial distribution
on longer time-scales. This separation of time-scales in the limit
$\xi^{-1}\rightarrow 0$ suggests the application of the multiple time-scale
analysis. This method replaces the physical distribution function $P(X,V;\tau)$
by an auxiliary function, $P(X,V;\tau_0,\tau_1,\tau_2,\dots)$, depending on
several time-scales. Accordingly, the time derivative in the physical evolution equation \gcall{4} is replaced by
the sum of time derivatives on each time-scale, 
$$
{\partial \over \partial \tau} \rightarrow {\partial \over \partial \tau_0}\
 +\ \xi_d^{-1}\ {\partial \over \partial \tau_1}\ +\ \xi_d^{-2}\ {\partial \over \partial \tau_2} + \dots
\tag{18}
$$
The auxiliary distribution function $P(X,V;\tau_0,\tau_1,\tau_2,\dots)$ is
then expanded in powers of the small parameter $\xi_d^{-1}$ and inserted
into the evolution equation, where terms of the same order are identified.
The physical solution of the system is eventually obtained by restricting
the different time variables to the so-called ``physical line''
$$
\tau_0 = \tau~;~ \tau_1=\xi_d^{-1}~\tau~;~\tau_2=\xi_d^{-2}~\tau~;~\dots
\tag{19}
$$
so that 
$$
P(X,V;\tau)=P^{(0)}(X,V;\tau,\xi_d^{-1}~\tau,\xi_d^{-2}~\tau,\dots)~+~
\xi_d^{-1}~P^{(1)}(X,V;\tau,\xi_d^{-1}~\tau,\xi_d^{-2}~\tau,\dots)~+~\dots
\tag{20}
$$
will be the solution of the Kramers equation \gcall{4}.
Equation \gcall{19} indicates that the dependence of the distribution
function on $\tau_n$ characterizes the evolution on the time-scale 
$\tau\sim \xi_d^{n}$ ($n=0,1,2,\dots$).

The crucial difference with the standard perturbation method is that
outside the physical line \gcall{19}, the auxiliary distribution
function has no physical meaning. Then appropriate boundary conditions
can be imposed to require the expansion to be uniform in the small 
parameter $\xi_d^{-1}$. This freedom will be used to
eliminate secular divergences.

The identification of different powers of $\xi_d^{-1}$ in the Kramers
equation gives the following relations
$$
\eqalign{
&\L_{FP}\ P^{(0)}\ = \ 0 \cr
&\L_{FP}\ P^{(1)}\ = \ \Biggl[{\partial \over \partial \tau_0}+V{\partial\over\partial X}
 +F(X){\partial \over \partial V}\Biggr] P^{(0)} \cr
&\L_{FP}\ P^{(2)}\ = \ \Biggl[{\partial \over \partial \tau_0}+V{\partial\over\partial X}
 +F(X){\partial \over \partial V}\Biggr] P^{(1)} + {\partial \over \partial \tau_1} P^{(0)} \cr
& \dots \cr
}
\tag{21}
$$

The zeroth order equation still imposes a maxwellian velocity distribution
$$
P^{(0)}(X,V;\tau_0,\tau_1,\dots)=\Phi (X;\tau_0,\tau_1,\dots) e^{-1/2 V^2}
\tag{22}
$$
where the function $\Phi$ now depends on all time variables.

The first order term thus
obeys to the following equation 
$$
\L_{FP}\ P^{(1)}\ = \ {\partial \Phi \over \partial \tau_0} e^{-1/2 V^2}+ V\left\{{\partial \Phi \over\partial X}
 -F\ \Phi \right\} e^{-1/2 V^2}
\tag{21b}
$$
The solubility condition then requires the function $\Phi$ to be
independent of the $\tau_0$ time-scale
$$
{\partial \Phi \over \partial \tau_0}=0
\tag{22b}
$$
and the first correction for the distribution function is now given by
$$
\eqalign{
P^{(1)}(X,V;\tau_0,\tau_1,\dots)=&-V\left\{{\partial \Phi(X;\tau_1,\dots) \over\partial X}
 -F\ \Phi (X;\tau_1,\dots)\right\} e^{-1/2 V^2}\cr
&+\Psi(X;\tau_0,\tau_1,\dots) e^{-1/2 V^2} \cr
}
\tag{23}
$$
The equation for $P^{(2)}$ becomes
$$
\eqalign{
\L_{FP}\ P^{(2)}\ =\ &\left\{ {\partial \Psi \over \partial \tau_0} 
+{\partial \Phi \over \partial \tau_1}
- {\partial \over \partial X } \left( {\partial \Phi \over\partial X}
 -F\ \Phi\right) \right\} e^{-1/2 V^2} + \left\{{\partial \Psi \over\partial X} -F\ \Psi \right\} V\ e^{-1/2 V^2} \cr
&-\left\{ 
F\left( {\partial \Phi \over\partial X}
 -F\ \Phi \right) -{\partial \over \partial X } \left( {\partial \Phi \over\partial X}
 -F\ \Phi\right) 
\right\} (1-V^2)\ e^{-1/2 V^2}
}
\tag{24}
$$
The solubility condition then requires
$$
{\partial \Psi \over \partial \tau_0} = - \left({\partial \Phi \over \partial \tau_1}
- {\partial \over \partial X } \left( {\partial \Phi \over\partial X}
 -F\ \Phi\right)  \right)
\tag{25}
$$
The r.h.s. of this equation does
not depend on $\tau_0$ (see eq. \gcall{22b}), and one must impose
the condition
$$
{\partial \Psi \over \partial \tau_0} = 0
\tag{26}
$$
to eliminate the secular divergence as $\tau_0$ grows to infinity.
This leads to a closed equation for $\Phi$
$$
{\partial \Phi \over \partial \tau_1}
- {\partial \over \partial X } \left( {\partial \Phi \over\partial X}
 -F\ \Phi\right) = 0 
\tag{27}
$$
This result has to be compared with eq. \gcall{13} of the standard Hilbert
procedure, where the time-dependence of the density was only
contained in the first order correction (through $\Psi$, see eq. \gcall{15}). A 
correct expansion thus modifies the perturbative scheme and leads to
different couplings between the successive corrections to the probability distribution.

The physically relevant equation for $\rho$ is then obtained by restricting
the different variables $\tau_i$ to the physical line \gcall{19}. The
time derivative of $\rho$ can now be written
$$
\eqalign{
{\partial \over {\partial \tau}} \rho(X,\tau)=&\sqrt{2\pi} 
{\partial \over {\partial \tau_0 }}\left[ \Phi(X;\tau_1,\dots) + \xi_d^{-1} \Psi(X;\tau_1,\dots)+ \O (\xi_d^{-2}) \right]_{\vert \tau_0=\tau,\tau_1=\tau/\xi_d,\dots} \cr
&+ {1\over \xi_d} \sqrt{2\pi} 
{\partial \over {\partial \tau_1 }}\left[ \Phi(X;\tau_1,\dots) + \xi_d^{-1} \Psi(X;\tau_1,\dots)+ \O (\xi_d^{-2}) \right]_{\vert \tau_0=\tau,\tau_1=\tau/\xi_d,\dots}  \cr
&= {1\over \xi_d} \sqrt{2\pi} 
{\partial \over {\partial \tau_1 }} \Phi(X;\tau_1,\dots)_{\vert \tau_0=\tau,\tau_1=\tau/\xi_d,\dots} + \O (\xi_d^{-2}) \cr
}
\tag{28}
$$
Using \gcall{27} and \gcall{15}, we eventually recover the (dimensionless) Smoluchowski equation $$
{\partial \rho(X,\tau) \over \partial \tau} 
= {1 \over \xi_d} {\partial \over \partial X } \left( {\partial \rho(X,\tau) \over\partial X}
 -F(X)\ \rho(X,\tau)\right) + \O (\xi_d^{-2})
\tag{29}
$$

\noindent
{\bf IV. DISCUSSION}
\taghead{IV.}

Let us make a few remarks on the present derivation. 

First, as mentionned in the introduction, the Smoluchowski equation
governs the spatial evolution of the Brownian particle. This evolution
is then expected to occur on a diffusion time scale $t_D \sim \ell^2/D$,
where $D=k_BT/M\xi$ is the diffusion coefficient, or in dimensionless
variables, $\tau_D = {v_T\over \ell} t_D \sim \xi_d$. This feature
emerges in a natural way from the multiple time scale analysis. Indeed,
according to equation \gcall{27}, we have shown that the spatial density probability evolves on the $\tau_1$ time scale, which precisely characterizes  the evolution of the system on
time scale $\tau \sim \xi_d \sim \tau_D$ (see discussion after equation \gcall{20}) !

As indicated above, the $t\gg \xi^{-1}$ limit of $P^{(0)}$ is just
the Maxwellian, which can be interpreted here as the ``local equilibrium 
approximation'' of the Fokker-Planck operator, $\L_{FP}$. As shown in eq. \gcall{22b}, this solution does not evolve on the time-scale $\tau_0$,
where this approximated form for $P^{(0)}$ remains valid.
This property confirms the relevance and the deep physical meaning of the local equilibrium approximation, 
as a state of the system which exists over a large time-scale.

Finally, let us remark that the formal analogy between the Kramers
equation and the Boltzmann's equation can serve as an introduction 
to perturbative expansion methods in Kinetic Theory. It
should be noted that the multiple-time scale method is indeed very close in spirit
to the Chapman-Enskog procedure. Indeed, the latter
introduces
an ad-hoc formal expansion of the time derivative in powers of the small
parameter $\epsilon$ 
$$
{\partial f \over \partial t}=\sum_{n=0}^{\infty} \epsilon^n 
{\partial^{(n)} f \over \partial t}
\tag{30}
$$ 
(in Kinetic Theory, the Knudsen number is the small parameter).
This is a subtle point of the procedure, which stems from the expansion 
in powers of $\epsilon$ of the functional dependence of
the distribution function 
on the ``hydrodynamic'' fields \refto{Titulaer78,Cercignani}. The explicit expressions for the operators $\partial^{(n)} / \partial t$ are computed
in the course of the calculation thanks to solubility conditions.
Equation \gcall{30} is formally equivalent  
to the multiple time-scale expansion of the time-derivative \gcall{18}. In this sense, the application of the
operator $\partial^{(n)} / \partial t$ of the Chapman-Enskog method
characterizes the dynamical evolution of the system on the time-scale
$t\sim \epsilon^{-n}$. For the purpose of a pedagogical approach, the multiple time-scale approach seems however more natural to us, since it gives
a clear understanding of each term of the expansion. Moreover it explicitly
shows the crucial importance of $\epsilon$ dependent corrections in the
time derivative \gcall{30} to obtain a consistent expansion. Absence 
of such terms would inevitably lead to inconsistencies identical to those
occuring in the naive expansion of the Kramers equation.

To conclude, the study of the Kramers equation in the high friction limit gives
a pedagogical and instructive presentation of the multiple time-scale method.
Pedagogical, since it can be shown explicitly that the method avoids all the troubles of more naive perturbation expansions; instructive, since it 
allows a deep insight into the underlying physics. This technique has already been successfully applied to a variety of problems in physics \refto{Nayfeh}, such as the Lorentz model \refto{Jarek}, adiabatic systems \refto{Anderson}
or a 
microscopic (exact) approach to Brownian motion \refto{Bocquet} to cite some
recent work. For such
complicated situations, it describes {\it in a systematic way} the dynamical evolution of the system on each successive time-scale. This property constitutes
the real power of the method.

{\bf Acknowledgments:} 

The author aknowledges stimulating discussions with M.-L. Citerne.

\endpage

\references

\refis{VK} N.G. van Kampen, {\it Stochastic Processes in Physics and Chemistry}
(North Holland, Amsterdam, 1990), 6th ed.

\refis{DGM} S.R. de Groot and P. Mazur, {\it Non-Equilibrium Thermodynamics}
(Dover Publication, New-York, 1984), pp. 111-142.

\refis{Cercignani} For a complete discussion of the Hilbert and Chapman-Enskog methods, see reference \refto{Titulaer78} and C. Cercignani, {\it The Boltzmann Equation and its Applications} (Springer-Verlag, 1988), pp. 232-261.

\refis{Klein21} O. Klein, ``The statistical theory of suspensions and 
solutions'', Ark. Mat. Astron. och Fysik {\bf 16}(5), 1-52 (1921).

\refis{Kramers40} H.A. Kramers, ``Brownian motion in a field of force and the diffusion model of chemical reactions'', Physica {\bf 7}, 284-304 (1940).

\refis{Smolu16} M. von Smoluchowski, ``Brownian molecular movement under the action of external forces and its connection with the generalized 
diffusion equation'', Ann. Physik {\bf 48}, 1103-1112 (1916).

\refis{Wilemski76} G. Wilemski, ``On the derivation of Smoluchowski equations
with corrections in the classical theory of Brownian motion'', J. Stat. Phys. {\bf 14}, 153-169 (1976).

\refis{Titulaer78} U.M. Titulaer, ``A systematic solution procedure for the 
Fokker-Planck equation of a Brownian particle in the high-friction case'',
Physica {\bf 91A}, 321-344 (1978); ``The Chapman-Enskog procedure as a form
of degenerate perturbation theory'', Physica {\bf 100A},
234-250 and ``Corrections to the Smoluchowski equation in the presence
of hydrodynamic interactions'', Physica {\bf 100A}, 251-265 (1980).

\refis{Jarek} J. Piasecki, ``Time scales in the dynamics of the Lorentz
electron gas'', Am. J. Phys. {\bf 61}, 718-722 (1993).

\refis{Bocquet} L. Bocquet, J. Piasecki, J.-P. Hansen, ``On the Brownian motion of a massive sphere in a hard-sphere fluid. I. Multiple-time-scale analysis
and microscopic expression for the friction coefficient'', J. Stat. Phys. 
{\bf 76}, 505-526 (1994).

\refis{Nayfeh} for further details and applications of the multiple time-scale
method, see A. Nayfeh, {\it Pertubation Methods} Wiley, New-York, 1973), pp.
228-307.

\refis{Anderson} J. L. Anderson, ``Multiple time scale methods for adiabatic
systems'', Am. J. Phys. {\bf 60}, 923-927 (1992).

\endreferences

\bye